\documentclass[journal]{IEEEtran}
\ifCLASSINFOpdf
\else
\fi
\usepackage{amsfonts,amsmath,amssymb}
\usepackage{mathtools}
\usepackage{mdwmath}
\usepackage{cuted}
\usepackage{mdwtab}
\usepackage{amsthm}           % for designing theorems
\usepackage{bm}               % provides bold math \bm{} command (vectors)
\usepackage{units}            % for nicely typset units \unit[10]{kHz}
\usepackage{graphicx}
\usepackage{epstopdf}
\usepackage[linesnumbered,ruled,vlined]{algorithm2e}

\usepackage{url}
\usepackage{paralist}
\usepackage{color}
\usepackage{authblk}
\usepackage{cleveref}
\usepackage{float}
\usepackage[caption = false]{subfig}
\newcommand{\Fig}[1]{Fig.~\ref{fig:#1}}

\newcommand{\Eq}[1]{Eq.~(\ref{eq:#1})}

\newcommand{\Lc}{\mathcal{L}}

\newcommand{\Rs}{\mathbb{R}^2}

\newcommand{\Ed}{\mathbb{E}}

\newcommand{\Pd}{\mathbb{P}}
\newcommand{\sinr}{\mathrm{SINR}}

\newcommand{\dr}{\mathrm{d}}

\newcommand{\Ru}{v(\gamma,r_{1})}
\newcommand{\Rl}{w(\gamma,r_{1})}

\newcommand{\Upper}{u}
\newcommand{\Lower}{l}


\DeclarePairedDelimiter\floor{\lfloor}{\rfloor}

\begin{document}
\title{Backhaul For Low-Altitude UAVs in Urban Environments}

\author{Boris Galkin,
        Jacek~Kibi\l{}da,
        and~Luiz~A. DaSilva
%\thanks{This material is based upon works supported by the Science Foundation
%Ireland under Grants No. 10/IN.1/I3007 and 14/US/I3110. B. Galkin, J. Kibi\l{}da, and L. DaSilva are with %CONNECT, Trinity College Dublin, Ireland, email:  \{galkinb,kibildj,dasilval\}@tcd.ie.}% <-this % stops a space
}

\affil{CONNECT- Trinity College Dublin, Ireland \\
\textit{E-mail: \{galkinb,kibildj,dasilval\}@tcd.ie}}

\maketitle

\begin{abstract}
Unmanned Aerial Vehicles (UAVs) acting as access points in cellular networks require wireless backhauls to the core network. In this paper we employ stochastic geometry to carry out an analysis of the UAV backhaul performance that can be achieved with a network of dedicated ground stations. We provide analytical expressions for the probability of successfully establishing a backhaul and the expected data rate over the backhaul link, given either an LTE or a millimeter-wave backhaul. We demonstrate that increasing the density of the ground station network gives diminishing returns on the performance of the UAV backhaul, and that for an LTE backhaul the ground stations can benefit from being colocated with an existing base station network. 
\end{abstract}


% \fi
\IEEEpeerreviewmaketitle

\section{Introduction}
Access points mounted on unmanned aerial vehicles (UAVs) are being proposed as a potential solution to data demand and congestion issues that are expected to arise in next-generation wireless networks \cite{Zeng_20162}. The benefit of these UAV networks compared to existing static infrastructure is the flexibility of their deployment, which allows them to configure themselves around user hotspots or gaps in coverage. These UAV networks additionally differ from existing infrastructure in how they connect into the core network; whereas fixed infrastructure can make use of wired backhaul links, often fiber optics, the UAVs must use dedicated wireless links for a backhaul. These links may be established using millimeter-waves \cite{Xiao_2016}, free-space optical channels \cite{Alzenad_2016} or using sub-6GHz technologies such as LTE. Emerging research in the wireless community suggests that existing LTE base stations (BSs) that are intended for ground user service are capable of also providing backhaul for UAV networks \cite{Lin_2017}. However, we expect that as UAV networks become more widespread and play a greater role in serving the end user network, operators will opt to deploy dedicated backhaul infrastructure, referred to as ground stations (GSs), to support these networks. UAVs have very different performance requirements and behave very differently to the typical user equipment; it follows that GS networks designed to serve UAVs will require different deployment strategies to BS networks designed to serve user equipment. 

One of the first works to consider low-altitude UAVs operating alongside terrestrial BS networks is \cite{Rohde_2012}. The authors simulate a network outage scenario where UAVs substitute offline BSs, with the UAVs wirelessly backhauling into adjacent, functioning BSs. The authors assume a hexagonal grid of BSs and demonstrate achieveable data rates in the downlink as a function of UAV distance to their backhaul. The works in \cite{Zeng_2016} and \cite{ZhangZeng_2017} approach the problem of UAV position optimisation by considering a single end user, UAV and backhaul in isolation, with the UAV adjusting its position to optimise its end user and backhaul links.
Stochastic geometry has also started to see use as a tool for modelling and analysing UAV-BS wireless links. In our previous work in \cite{Galkin_20172} (which extends our work in \cite{Galkin_2017}) we investigated the performance of a UAV network when the UAVs opportunistically backhaul through LTE BSs designed to serve end users on the ground. We considered both the backhaul link as well as the link between the UAV and the end user, and we demonstrated that while the two different links have optimal performance at differing UAV heights it is possible to maximise the performance of both links by allowing UAVs to adjust their individual heights. A similar analysis of the UAV-BS link was carried out in \cite{MahdiAzari_20172} where the authors consider the performance of UAV-BS links in different environments, given down-tilted BS antennas and omnidirectional UAV antennas. The authors conclude that the UAV-BS link is vulnerable to interference due to the high LOS probability, and suggest that both UAVs and BSs reduce their heights above ground to improve performance.

While a limited number of results have been published on the interaction between UAV networks and existing terrestrial BS networks, to date there is a lack of work on the topic of analysing the performance of a dedicated GS network that is deployed for providing backhaul to UAV networks. Our contribution in this paper is to use stochastic geometry to analyse the backhaul performance of GSs serving a low-altitude UAV network in an urban environment. We consider the use of both 2GHz LTE and millimeter-wave technology to support the backhaul link, with the GSs using frequency bands that are orthogonal to the underlying BS networks. Our analysis takes into account directional antenna alignment in 3D space between a typical UAV in the network and its associated backhaul GS as well as other interfering GSs, and it captures the effect of generalised multipath fading as well as line-of-sight (LOS) blocking due to buildings in the environment. We explore the effect of GS density as well as their height above ground on the probability of a typical UAV being able to establish a backhaul, given different operating heights of the UAV. We initially assume that the GSs are deployed independently of existing terrestrial infrastructure; in the numerical results section we show that when using an LTE link for the backhaul the GS infrastructure can be colocated with existing BS sites to achieve good backhaul performance. The analysis in this work allows us to address basic network design questions that would be relevant to a network operator interested in providing dedicated backhaul infrastructure for a network of UAVs.

\begin{figure}[t!]
\centering
	\subfloat{\includegraphics[width=.40\textwidth]{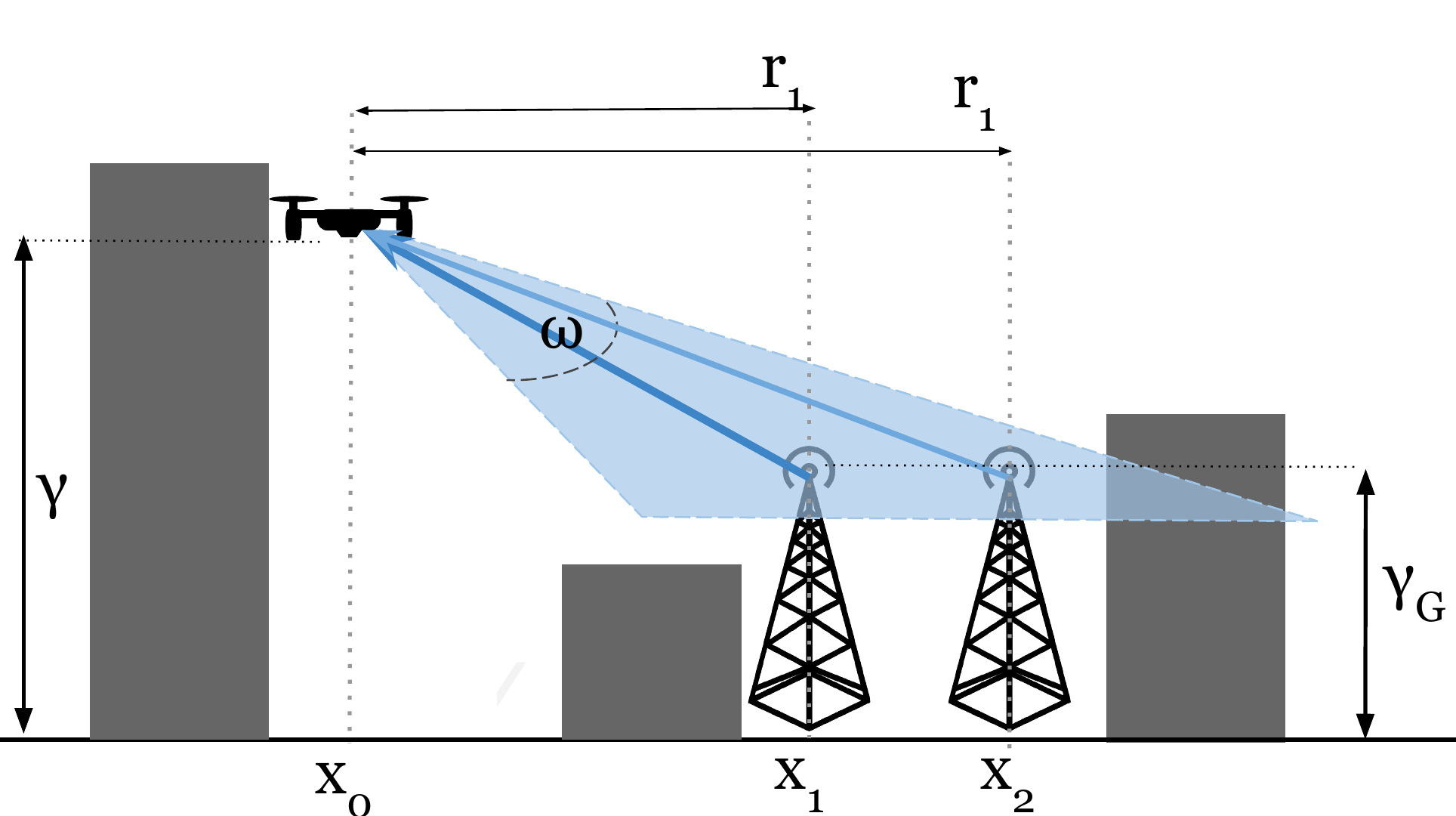}}\\
	\subfloat{\includegraphics[width=.40\textwidth]{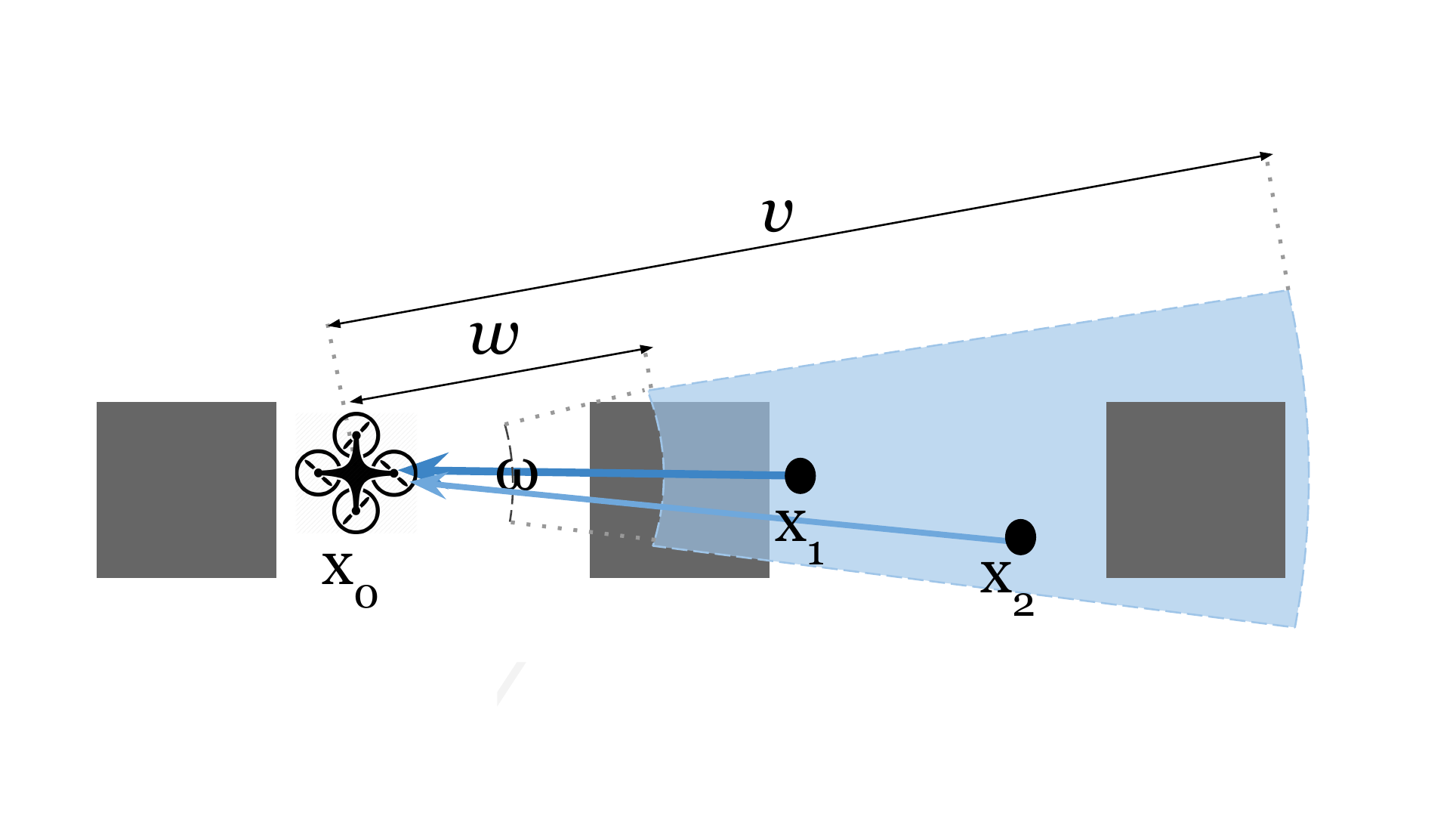}}
%	\vspace{-3mm}
	\caption{
	Side and top view showing a UAV in an urban environment at a height $\gamma$, positioned above $x_0$ with antenna beamwidth $\omega$. The UAV selects the nearest GS at $x_1$ for its backhaul and aligns its antenna with the GS location; the GS at $x_2$ falls inside the aligned area and potentially produces interference.
	\vspace{-5mm}
	}
	\label{fig:drone_network}
\end{figure}

\section{System Model}

In this section we set up a system model of a network of GSs providing backhaul to UAVs. We consider two technologies for the backhaul channel, LTE in the 2GHz band and millimeter-wave. We model the network of GSs as a Poisson point process (PPP) $\Phi = \{x_1 , x_2 , ...\} \subset \Rs$ of intensity $\lambda$ where elements $x_i\in \Rs$ represent the projections of the GS locations onto the $\Rs$ plane.  The GSs have a height $\gamma_{G}$ above ground. We consider a single reference UAV, positioned above the origin $x_0 = (0,0)$ at a height $\gamma$. Let $r_i = ||x_i||$ denote the horizontal distance between the GS $i$ and the reference UAV, and let $\phi_i = \tan^{-1}((\Delta \gamma)/r_i)$ denote the vertical angle, where $\Delta \gamma = \gamma - \gamma_{G}$. 

The UAV is equipped with a directional antenna for communicating with its associated backhaul GS. The antenna has a horizontal and vertical beamwidth $\omega$ and a rectangular radiation pattern; using the approximations (2-26) and (2-49) in \cite{Balanis_2005} and assuming perfect radiation efficiency the antenna gain can be expressed as $\eta(\omega) = 16\pi/(\omega^2)$ inside of the main lobe and $\eta(\omega)=0$ outside. The UAV selects the nearest GS as its serving GS; we denote this GS as $x_1$ and its distance to the UAV as $r_1$. The UAV orients itself to align its backhaul antenna towards $x_1$; the antenna radiation pattern illuminates an area we denote as $\mathcal{W} \subset \Rs$. This area takes the shape of a ring sector of arc angle equal to $\omega$ and major and minor radii $\Ru$ and $\Rl$, respectively, which are defined as

\begin{align}
\Ru = 
\begin{cases}
\frac{|\Delta \gamma|}{\tan(|\phi_{1}|-\omega/2)} \hspace{-2mm} &\text{if} \hspace{3mm} \omega/2 < |\phi_{1}| < \pi/2 - \omega/2 \\
\frac{|\Delta \gamma|}{\tan(\pi/2 -\omega)} \hspace{-2mm} &\text{if} \hspace{3mm} |\phi_{1}| > \pi/2 - \omega/2 \\
\infty &\text{otherwise}
\end{cases}
\end{align}

\begin{align}
\Rl = 
\begin{cases}
\frac{|\Delta \gamma|}{\tan(|\phi_{1}|+\omega/2)} \hspace{2mm} &\text{if} \hspace{3mm} |\phi_{1}| < \pi/2 - \omega/2  \\ 
0 &\text{otherwise}
\end{cases}
\end{align}

\noindent
where $|.|$ denotes absolute value. For the case where $\omega\geq \pi/2$ major radius $\Ru$ will always be infinite. We denote the set of GSs other than $x_1$ that fall inside the area $\mathcal{W}$ as $\Phi_{\mathcal{W}} = \{x_i \in \Phi \setminus \{x_1\} : x_i \in \mathcal{W}\}$. The GSs in the set $\Phi_{\mathcal{W}}$ are capable of causing interference for the reference UAV. Those GSs in the set $\Phi \setminus \Phi_{\mathcal{W}}$ are outside of the main lobe of the UAV antenna, thus they do not interfere with the backhaul link. Note that $\Phi_{\mathcal{W}}$ is a PPP with the same intensity $\lambda$.

The wireless channels between the reference UAV and the GSs will be affected by buildings, which form obstacles and break LOS links. We adopt the model in \cite{ITUR_2012}, which defines an urban environment as a collection of buildings arranged in a square grid. There are $\beta$ buildings per square kilometer, the fraction of area occupied by buildings to the total area is $\delta$, and each building has a height which is a Rayleigh-distributed random variable with scale parameter $\kappa$. The probability of the reference UAV having LOS to the GS $i$ is given in \cite{ITUR_2012} as 
\vspace{-1mm}
\begin{align}
&\Pd_{LOS}(r_{i}) = \nonumber \\
&\prod\limits_{n=0}^{\max(0,d_i-1)}\left(1-\exp\left(-\frac{\left(\max(
\gamma,\gamma_{G}) - \frac{(n+1/2)|\Delta \gamma|}{d_i}\right)^2}{2\kappa^2}\right)\right)
\label{eq:LOS}
\vspace{-1mm}
\end{align}
%Let us denote the distance to UAV $i$ as $r_i = ||x_i||$. The new set $\Phi_{\mathcal{W}}' = \{r_i \in \Rb\}$ is an inhomogeneous Poisson point pattern with intensity $\lambda'(r) = 2\pi\lambda r$. 

\noindent
where $d_i = \floor*{r_i\sqrt{\beta\delta}}$. We can express the SINR at the reference user as 

\begin{equation}
\sinr = \frac{p H_{t_{1}} \eta(\omega)\mu c (r_{1}^2+\Delta \gamma^2)^{-\alpha_{t_{1}}/2}}{I_{L} + I_{N}+\sigma^2}
\label{eq:SINR}
\end{equation}

\noindent
 where $p$ is GS transmit power, $H_{t_1}$ is the random multipath fading component, $t_{1} \in \{\text{L},\text{N}\}$ is an indicator variable which denotes whether the UAV has LOS or NLOS to its serving GS, $\mu$ is the serving GS antenna gain defined in the next subsections, $c$ is the near-field pathloss, $\sigma^2$ is the noise power, and $I_{L}$ and $I_{N}$ are the aggregate LOS and NLOS interference, respectively. Expressions for the noise power $\sigma^2$ and near-field pathloss $c$ are given in \cite{Elshaer_2016}.
 
 The backhaul data rate, in Mbits/sec, can be calculated from the SINR using the Shannon capacity bound
 
 \begin{equation}
 \mathcal{R} = b \log_2(1+\sinr)
 \end{equation}

\noindent
where $b$ denotes the bandwidth of the backhaul channel. We define a SINR threshold $\theta$ for the UAV backhaul link: if $\sinr<\theta$ this represents the UAV failing to establish a backhaul of the required channel quality and therefore being in an outage state.

\subsection{LTE Backhaul}
For the LTE backhaul case we assume that the GSs are equipped with tri-sector antennas similar to those already in use in terrestrial BSs, as this allows them to serve UAVs in any horizontal direction. For tractability we model the horizontal antenna gain $\mu_{h}$ of these antennas as having a constant value. The antennas are tilted up towards the sky, to model the behaviour of the antennas in the vertical plane we adopt the 3GPP model \cite{3GPP_2010}, such that

\begin{equation}
\mu_{v}(\phi_{i}) = 10^{-\min\left(12\left(\frac{\phi_{i} - \phi_T}{10}\right)^2, 20 \right)/10},
\end{equation}

\begin{equation}
\mu_{l}(\phi_{i}) = \max\left(\mu_{h}\mu_{v}(\phi_{i}),10^{-2.5}\right),
\end{equation}

\noindent
where $\mu_{v}(\phi_{i})$ is the vertical antenna gain, $\phi_T$ is the vertical uptilt angle of the GS antenna (in degrees) and $\mu_{l}(\phi_{i})$ is the total antenna gain.

\subsection{Millimeter Wave Backhaul}

For the millimeter-wave backhaul we assume each GS is equipped with an antenna array that uses beamforming to direct a directional beam towards the UAV to which it provides a backhaul link. We adopt a similar approach to modelling the GS antenna array as in \cite{Elshaer_2016} and \cite{Andrews_2017}. The GS antenna is modelled as having a single directional beam with a beamwidth of $\omega_G$ and a gain of $\mu_{m}$ inside the main lobe, and a gain of 0 outside. The reference UAV will always experience an antenna gain of $\mu_{m}$ from its serving GS. The beam patterns of the remaining GSs will appear to be pointed in random directions with respect to the reference UAV; as a result each interfering GS will have non-zero antenna gain to the reference UAV with a certain probability $\zeta$. 

\section{Analytical Results}
In this section we derive an analytical expression for the probability that the reference UAV will receive a signal from the GS network with an SINR above $\theta$, thereby establishing a backhaul. We refer to this as the backhaul probability. To derive an expression for the backhaul probability we need an expression for the conditional backhaul probability given the serving GS of the reference UAV has either LOS or NLOS to the reference UAV, and given it is located at a horizontal distance $r_1$ from the UAV. We then decondition this conditional backhaul probability with respect to the LOS probability of the serving GS as well as its horizontal distance. The LOS probability for a given horizontal distance $r_1$ is given in \Eq{LOS}. Given a PPP distribution of GSs the serving GS horizontal distance random variable $R_1$ is known to be Rayleigh-distributed with scale parameter $1/\sqrt{2\pi\lambda}$.

\subsection{Aggregate LOS \& NLOS Interference}
\textbf{LTE backhaul}
 The interferers will belong to the set $\Phi_{\mathcal{W}}$. We partition this set into two sets which contain the LOS and NLOS interfering GSs, denoted as $\Phi_{\mathcal{W}L} \subset \Phi_{\mathcal{W}}$ and $\Phi_{\mathcal{W}N} \subset \Phi_{\mathcal{W}}$, respectively. These two sets are inhomogeneous PPPs with intensity functions $\lambda_L(x) = \Pd_{LOS}(||x||)\lambda$ and $\lambda_N(x) =(1-\Pd_{LOS}(||x||))\lambda$. Note that we drop the index $i$ as the GS coordinates have the same distribution irrespective of their index values. For an LTE backhaul the aggregate LOS and NLOS interference is expressed as $I_{L} = \sum_{x\in\Phi_{\mathcal{W}L}} p H_{L} \eta(\omega)\mu_l(\phi) c (||x||^2+\Delta \gamma^2)^{-\alpha_{L}/2}$ and $I_{N} = \sum_{x\in\Phi_{\mathcal{W}N}} p H_{N} \eta(\omega)\mu_l(\phi) c (||x||^2+\Delta \gamma^2)^{-\alpha_{N}/2}$, recalling that $\phi = \tan^{-1}(\Delta \gamma/||x||)$.
 
 \textbf{millimeter-wave backhaul}
 As defined in the system model, the millimeter-wave interfering GSs will only create interference at the reference UAV if their directional beams happen to align with the UAV location, which occurs with probability $\zeta$. As a result of this $\Phi_{\mathcal{W}L}$ and $\Phi_{\mathcal{W}N}$ have intensity functions $\lambda_L(x) = \Pd_{LOS}(||x||)\zeta\lambda$ and $\lambda_N(x) =(1-\Pd_{LOS}(||x||))\zeta\lambda$. The aggregate LOS and NLOS interference is then expressed as $I_{L} = \sum_{x\in\Phi_{\mathcal{W}L}} p H_{L} \eta(\omega)\mu_m c (||x||^2+\Delta \gamma^2)^{-\alpha_{L}/2}$ and $I_{N} = \sum_{x\in\Phi_{\mathcal{W}N}} p H_{N} \eta(\omega)\mu_m c (||x||^2+\Delta \gamma^2)^{-\alpha_{N}/2}$.
 %As we are interested in the distances of the GSs to the UAV rather than their absolute coordinates in $\Rs$ in the following derivations we use the polar coordinates of the GS locations, such that $x_i$ = $(r_i,\rho_i)$, where $\rho_i$ is the angle relative to the x-axis. As a PPP is isotropic we align the x-axis with the line between the UAV at the origin and the backhaul GS $x_1$, so that $\rho_1 = 0$. We partition $\Phi_{\mathcal{W}}$ into two PPPs which contain the GSs that have LOS and NLOS channels to the UAV, denoted as $\Phi_{\mathcal{W}L} \subset \Rs$ and $\Phi_{\mathcal{W}N} \subset \Rs$, respectively, with intensity functions $\lambda_{L}(r) = \Pd_{LOS}(\gamma,\gamma_{GS},r) \lambda$ and $\lambda_{N}(r) = (1-\Pd_{LOS}(\gamma,\gamma_{GS},r)) \lambda$. Note that we drop the index $i$. In effect, the LOS probability function acts as a thinning function \cite{Haenggi_2013} which removes (thins) UAVs from the PPP $\Phi_{\mathcal{W}}$ with probability $(1-\Pd_{LOS}(\gamma,\gamma_{GS},r))$ to form $\Phi_{\mathcal{W}L}$ from the remaining UAVs and $\Phi_{\mathcal{W}N}$ from those that are thinned. The aggregate LOS and NLOS interference is then described as $I_L=\sum_{r\in\Phi_{\mathcal{W}L}}p\eta(\omega_{bs})\eta_{GS}(\phi) H_L (r^2+\Delta\gamma^2)^{-\alpha_L/2}$ and $I_N=\sum_{r\in\Phi_{\mathcal{W}N}}p\eta(\omega_{bs})\eta_{GS}(\phi)^{-\alpha_N/2}$.  

\subsection{Conditional Backhaul Probability}
%In this subsection we describe how the coverage probability for the UAV network is obtained as a function of the network design and wireless channel parameters.
\noindent
%Considering Nakagami-$m$ fading, the conditional coverage probability $\Pd(\sinr\geq \theta |R_1=r_1)$ is obtained following (21) in \cite{Chetlur_2017} as

%\begin{equation}
 %\Pd(\sinr\geq \theta |R_1=r_1) =  \\
% \sum\limits_{n=0}^{m_{t_1}-1}\frac{s_{t_1}^n}{n!} (-1)^n \frac{d^n %\Lc_{I}((p\eta(\omega))^{-1}s_{t_1})}{ds_{t_1}^n},
%\end{equation}

%\noindent
%where $s_{t_1}= m_{t_1}(\eta_{GS}(\phi_1)^{-1})\theta(r_1^2+\Delta\gamma^2)^{\alpha_{t_1}/2}$, $m_{t_1}$ is the Nakagami-$m$ fading term and $\Lc_{I}$ denotes the Laplace transform of the total interference. The LOS and NLOS interferers are distributed independently of one another; the proof of this is similar to the proof in \cite{Haenggi_2013} and is omitted here. Due to this, the Laplace transform above can be separated into a product of the Laplace transforms of the aggregate LOS and aggregate NLOS interference, along with the introduction of the noise-related term. This allows us to express the conditional coverage probability for an LOS serving UAV $\Pd(\sinr\geq \theta |R_1=r_1,t_1 = \text{L})$ as 

The expression for the backhaul probability, given serving GS distance $r_1$ and an LOS channel to the serving GS, was derived by us in \cite{Galkin_2017} as

\begin{align}
&\Pd(\sinr\geq \theta |R_1=r_1,t_1 = \text{L}) = \nonumber \\
&\sum\limits_{n=0}^{m_L-1}\frac{s_L^n}{n!} (-1)^n 
 \cdot \sum_{i_L+i_N+i_{\sigma}=n}\frac{n!}{i_L!i_N!i_{\sigma}!} \nonumber \\
 &\cdot(-(p\eta(\omega)c)^{-1}\sigma^2)^{i_{\sigma}}\exp(-(p\eta(\omega)c)^{-1}s_L\sigma^2) \nonumber \\
 &\cdot\frac{d^{i_L} \Lc_{I_{L}}((p \eta(\omega)c)^{-1}s_L)}{ds_L^{i_L}} \frac{d^{i_N}\Lc_{I_{N}}((p  \eta(\omega)c)^{-1}s_L)}{ds_L^{i_N}},
\label{eq:condProb3}
\end{align}

\noindent
where $s_{L}= m_{L}\theta \mu^{-1}(r_1^2+\Delta\gamma^2)^{\alpha_{L}/2}$, $m_L$ is the Nakagami-$m$ fading term for a LOS channel, $\Lc_{I_{L}}$ and $\Lc_{I_{N}}$ are the Laplace transforms of the aggregate LOS and NLOS interference, respectively, and the second sum is over all the combinations of non-negative integers $i_L,i_N$ and $i_{\sigma}$ that add up to $n$. $\mu$ takes the value of either $\mu_l(\phi_1)$ or $\mu_m$ depending on whether we are considering LTE or millimeter-wave backhaul. The conditional backhaul probability given an NLOS serving GS $\Pd(\sinr\geq \theta |R_1=r_1,t_1 = \text{N})$ is calculated as in \Eq{condProb3} with $m_N$, $\alpha_N$ and $s_N$ replacing $m_L$, $\alpha_L$ and $s_L$. 

\subsection{Laplace Transform of Aggregate Interference}
\textbf{LTE backhaul}
 The Laplace transform of the aggregate LOS interference $\Lc_{I_{L}}((p\eta(\omega)c)^{-1}s_L)$ given an LOS serving GS is expressed as
\begin{align}
&\Ed\left[\exp\bigg(-(p\eta(\omega)c)^{-1}s_L I_{L}\bigg)\right] \nonumber \\
&=\Ed_{\Phi_{\mathcal{W}L}}\bigg[\prod_{x\in\Phi_{\mathcal{W}L}}\hspace{-3mm}\Ed_{H_L} \left[\exp\Big(-H_L g(||x||,s_L,\alpha_L)\Big)\right]\bigg]  \nonumber 
\end{align}
\begin{align}
&\overset{(a)}{=}\Ed_{\Phi_{\mathcal{W}L}}\left[\prod_{x\in\Phi_{\mathcal{W}L}}\left(\frac{m_L}{g(||x||,s_L,\alpha_L)+m_L}\right)^{m_L}\right]  \nonumber \\
%\overset{(b)}{=} \exp\Bigg(-\int\limits_{\mathcal{W}} \left(1-g(r,s_L,m_L,\alpha_L)\right) \lambda'_{L}(r) \dr x\Bigg)
&\overset{(b)}{=} \exp\Bigg(\hspace{-1mm}-\int\limits_{\mathcal{W}}\Bigg(1- \left(\frac{m_L}{g(||x||,s_L,\alpha_L)+m_L}\right)^{m_L}\hspace{-1mm}\Bigg) \lambda_{L}(x) \dr x \Bigg) \nonumber \\
&\overset{(c)}{=} \exp\hspace{-1mm}\Bigg(\hspace{-1mm}-\hspace{-1mm}\lambda \omega \hspace{-4mm} \int\limits_{r_1}^{\Ru}\hspace{-2mm}\Bigg(\hspace{-1mm}1- \hspace{-1mm}\left(\frac{m_L}{g(r,s_L,\alpha_L)+m_L}\right)^{m_L}\hspace{-1mm}\Bigg) \Pd_{LOS}(r)r \dr r \hspace{-1mm}\Bigg) \nonumber \\
\label{eq:laplace}
\end{align}

\noindent
where 
\begin{equation}
 g(||x||,s_L,\alpha_L) = s_L\mu_{l}(\phi)(||x||^2+\Delta\gamma^2)^{-\alpha_L/2}   \nonumber,
\end{equation}

\noindent
 $(a)$ comes from Nakagami-$m$ fading having a gamma distribution, $(b)$ comes from the probability generating functional of the PPP \cite{Haenggi_2013}, $(c)$ comes from switching to polar coordinates where $r = ||x||$ and $\lambda_{L}(x) = \Pd_{LOS}(||x||) \lambda$. Note that the Laplace transform for the NLOS interferers $\Lc_{I_N}((p \eta(\omega) c)^{-1}s_L)$ is solved by simply substituting $\lambda_{L}(x)$ with $\lambda_{N}(x)$, $m_L$ with $m_N$ and $g(r,s_L,\alpha_L)$ with $g(r,s_L,\alpha_N)$ in \Eq{laplace} and solving as shown. The above integration is for the case when the serving GS is LOS; if the serving GS is NLOS we substitute $s_L$ with $s_N$ as defined in the previous subsection.
 
 \textbf{millimeter-wave backhaul} The Laplace transform of the LOS interferers for a millimeter-wave backhaul is derived as in \Eq{laplace}, with the intensity $\lambda$ being multiplied by $\zeta$ (as explained in the previous subsection), and with $\mu_{l}(\phi)$ being replaced with $\mu_{m}$. Note that, unlike $\mu_{l}(\phi)$, $\mu_{m}$ is a constant value with respect to $r$; as a result of this it is possible to solve the integral in \Eq{laplace} for the case of millimeter-wave backhaul. We begin by recognising that $\Pd_{LOS}(r)$ is a step function. We use this fact to separate the integral above into a sum of weighted integrals, resulting in the following expression
 
 %\begin{equation}
%\omega\lambda\sum\limits_{j=\floor*{\Rl\sqrt{\beta\delta}}}^{\floor*{\Ru\sqrt{\beta\delta}}} %\Pd_{LOS}(l)\int\limits_{l}^{u}(1-g(r,s_L,m_L,\alpha_L))r\dr r 
%\label{eq:laplaceSum}
% \end{equation}

\begin{equation}
\omega\zeta\lambda\hspace{-5mm}\sum\limits_{j=\floor*{r_1\sqrt{\beta\delta}}}^{\floor*{\Ru\sqrt{\beta\delta}}} \hspace{-5mm}\Pd_{LOS}(l)\int\limits_{l}^{u}\Bigg(1- \left(\frac{m_L}{g(r,s_L,\alpha_L)+m_L}\right)^{m_L}\Bigg) r \dr r
\label{eq:laplaceSum}
\end{equation}

\noindent
where $l = \max(r_1,j/\sqrt{\beta\delta})$ and $u = \min(\Ru,(j+1)/\sqrt{\beta\delta})$. Using the derivations (17) in \cite{Galkin_2017}

\begin{align}
&\int\limits_{l}^{u}\hspace{-1mm}\Bigg(1- \hspace{-1mm}\left(\frac{m_L}{g(r,s_L,\alpha_L)+m_L}\right)^{m_L}\hspace{-1mm}\Bigg) r \dr r = \frac{1}{2}\sum\limits_{k=1}^{m_L}\hspace{-1mm}\binom{m_L}{k}(-1)^{k+1}\nonumber \\
&\cdot\bigg((\Upper^2+\Delta\gamma^2)\mbox{$_2$F$_1$}\Big(k,\frac{2}{\alpha_L};1+\frac{2}{\alpha_L};  -\frac{m_L(\Upper^2+\Delta\gamma^2)^{\alpha_L/2}}{\mu_{m} s_L}\Big) \nonumber\\
&-(\Lower^2+\Delta\gamma^2)\mbox{$_2$F$_1$}\Big(k,\frac{2}{\alpha_L};1+\frac{2}{\alpha_L};-\frac{m_L(\Lower^2+\Delta\gamma^2)^{\alpha_L/2}}{\mu_{m} s_L}\Big)\bigg),
\label{eq:laplace_final}
\end{align}

\noindent
Inserting this solution into \Eq{laplaceSum} we obtain an expression for the Laplace transform of the LOS interferers \Eq{laplace}.

\subsection{Backhaul Probability and Expected Rate}

%\begin{proposition}
To obtain the overall backhaul probability for the reference UAV in the network we decondition the conditional backhaul probability as defined in the previous subsection with respect to the indicator variable $t_1$ by multiplying by the LOS probability function \Eq{LOS}. We then decondition with respect to the horizontal distance random variable $R_1$ via integration.
%\end{proposition}

\begin{multline}
\Pd(\sinr\geq \theta) = 
\int\limits_{0}^{\infty}\bigg(\Pd(\sinr\geq \theta |R_1=r_1,t_1 = \text{L})\Pd_{LOS}(r_1) \\
+\Pd(\sinr\geq \theta |R_1=r_1,t_1 = \text{N})(1-\Pd_{LOS}(r_1))\bigg)f_{R_1}(r_1) \dr r_1 .
 \label{eq:pcov_final}
\end{multline}

The expected rate for the backhaul can be calculated using the backhaul probability as 

\begin{equation}
\Ed[\mathcal{R}] = b\int\limits_{0}^{\infty} \Pd(\sinr\geq 2^{\theta}-1) \dr \theta.
\end{equation}

 \section{Numerical Results}

In this section we explore the trade-offs that occur between the parameters of the GS network and the resulting backhaul probability of the reference UAV. We generate our results using the analytical expressions given in the previous section and validate them via Monte Carlo (MC) trials. In the following figures solid lines denote the results obtained via the mathematical analysis and markers denote results obtained via MC trials. Unless stated otherwise the parameters used for the numerical results are taken from Table \ref{tab:table}.

\begin{table}[b!]
\vspace{-3mm}
\begin{center}
\caption{Numerical Result Parameters}
\begin{tabular}{ |c|c|c| } 
 \hline
 Parameter & LTE Backhaul & mmWave Backhaul  \\ 
 \hline
 Carrier Freq & \unit[2]{GHz} & \unit[73]{GHz} \\
 Bandwidth & \unit[20]{MHz} & \unit[1000]{MHz} \\
 $\omega$ & \unit[30]{deg} & \unit[10]{deg} \\
 $\alpha_L$ & 2.1 & 2\\
 $\alpha_N$ & 4 & 3.5 \cite{Ghosh_2014}\\
 $m_L$ & 1 & 3\\
 $m_N$ & 1 & 1\\
 $p$ & \unit[40]{W} & \unit[10]{W} \cite{Semiari_2017}\\
 $c$ & \unit[-38.4]{dB} & \unit[-69.7]{dB} \\
 $\mu_{h}$ & \unit[-5]{dB} & N/A \\
 $\mu_{m}$ & N/A & \unit[32]{dB} \\
 $\zeta$ & N/A & 0 \\
 $\theta$ & \unit[10]{dB} & \unit[10]{dB} \\
 $\sigma^2$ & \unit[$8\cdot10^{-13}$]{W} &\unit[$4\cdot10^{-11}$]{W} \\
 $\phi_T$ & $\tan^{-1}(\Delta\gamma/\Ed [R_1] )$ & N/A \\ 
 $\omega_G$ & N/A & \unit[10]{deg} \\
  $\gamma_{G}$ &  \unit[30]{m} & \unit[30]{m}\\
 $\beta$ & \unit[300]{$/\text{km}^2$} & \unit[300]{$/\text{km}^2$}\\
 $\delta$ & 0.5 & 0.5\\
 $\kappa$ & \unit[20]{m} & \unit[20]{m} \\
 \hline
\end{tabular}
 \label{tab:table}
\end{center}
\end{table}

\begin{figure}[b!]

\centering
	\includegraphics[width=.45\textwidth]{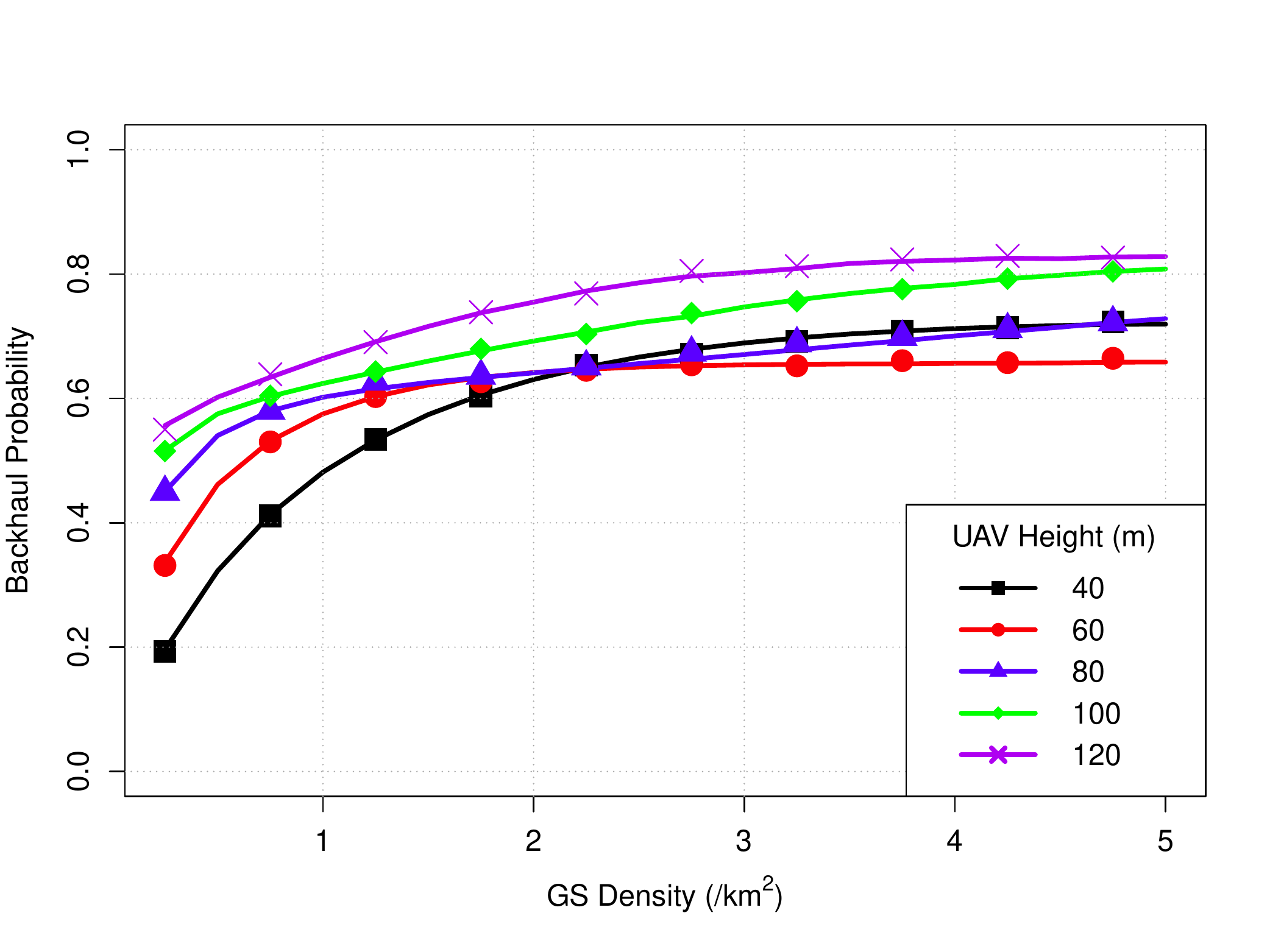}
    \vspace{-5mm}
	\caption{
	Backhaul probability for an LTE backhaul as a function of the GS density.
	}
	\label{fig:BackhaulDensity}
	\vspace{-3mm}
\end{figure}

\begin{figure}[t!]

\centering
	\includegraphics[width=.45\textwidth]{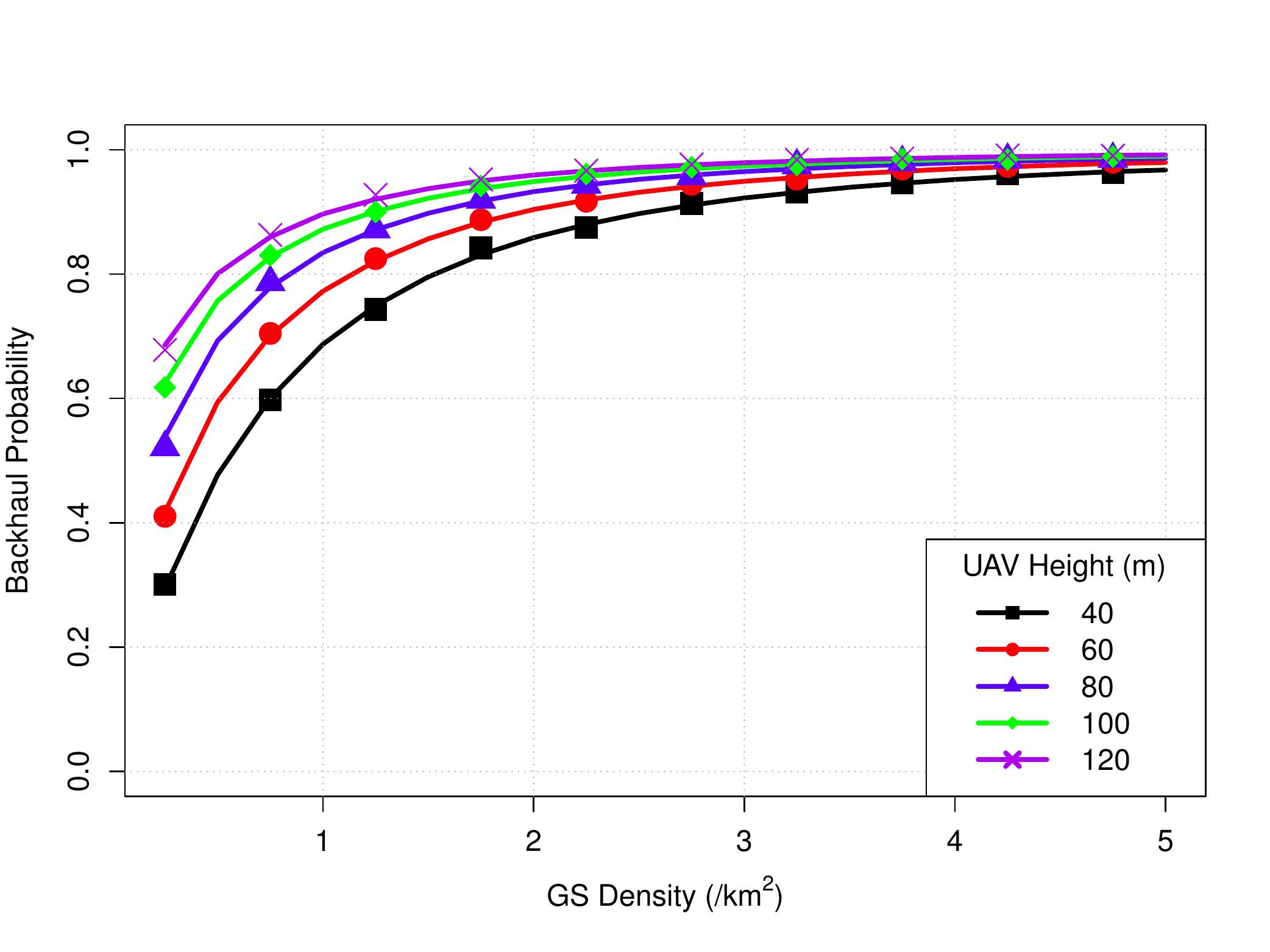}
    \vspace{-5mm}
	\caption{
	Backhaul probability for a millimeter-wave backhaul as a function of the GS density.
	}
	\label{fig:mmwaveDensity}
	\vspace{-3mm}
\end{figure}

In \Fig{BackhaulDensity} and \Fig{mmwaveDensity} we demonstrate how increasing the density of the GSs improves the backhaul probability of the UAV, for the LTE and the millimeter-wave backhaul cases, respectively. For all cases the backhaul probability increases asymptotically, with more GSs giving diminishing returns on the improved performance. Note that we set the millimeter-wave interference probability parameter $\zeta$ to zero, to reflect the fact the millimeter-wave antennas have very narrow beamwidths and therefore have an extremely low probability of alignment occuring by chance. As a result of this the millimeter-wave signal is noise limited, with a resulting higher backhaul probability for the UAV compared to the interference-limited LTE signals. We consider the upper limit for the GS density to be \unit[5]{$/\text{km}^2$}, which corresponds to the density of a terrestrial BS network in an urban environment \cite{3GPP_2010}. These results suggest that good UAV backhaul probability can be achieved when the density of GSs is only a fraction of the density of the existing BS network; if the GSs are to be colocated with the BS sites then this result demonstrates that a network operator only has to upgrade a fraction of the existing BS network to be able to provide a backhaul for a network of UAVs.

\begin{figure}[t!]
\centering
	\includegraphics[width=.45\textwidth]{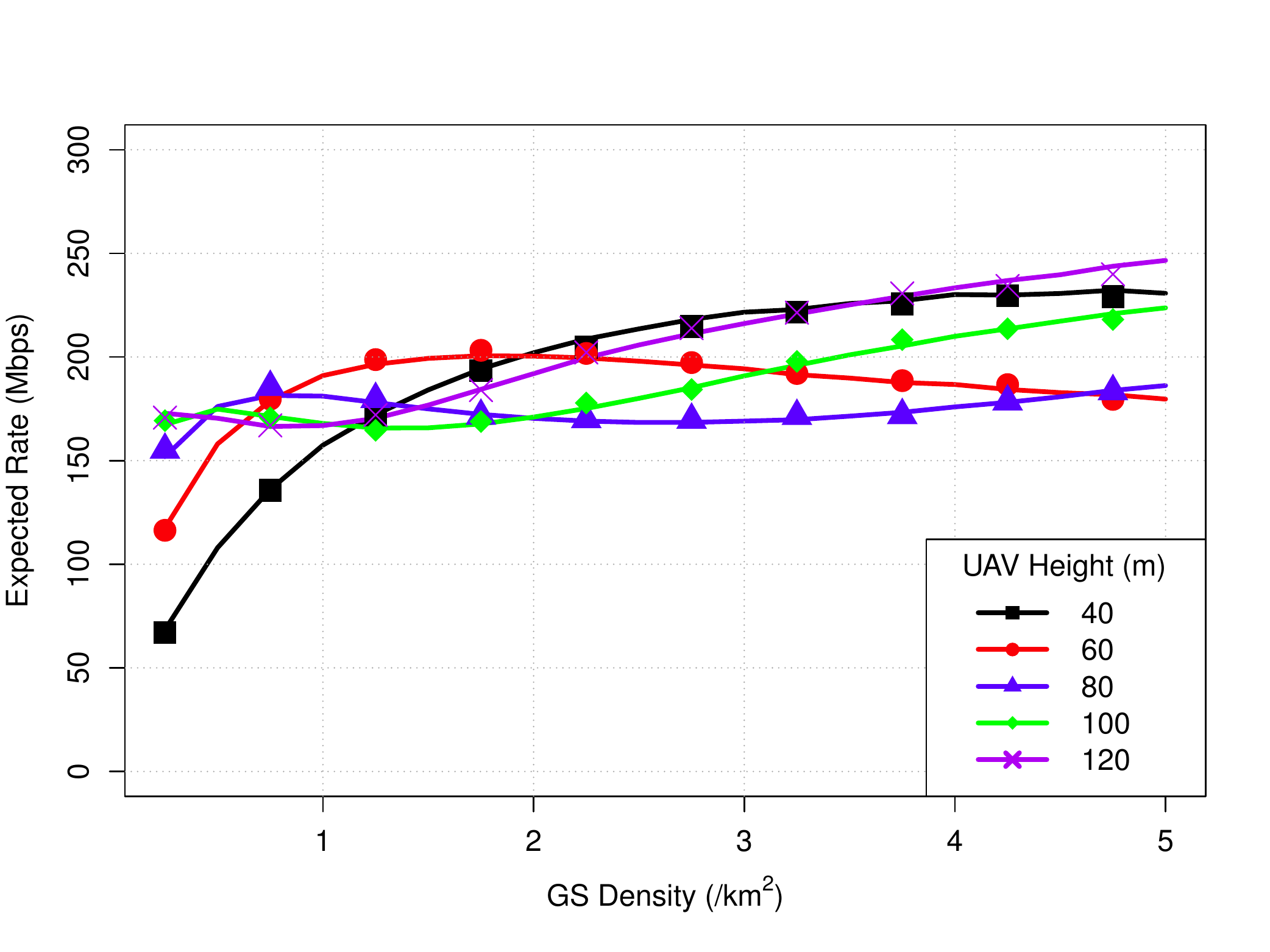}
    \vspace{-5mm}
	\caption{
	Expected data rate of a backhaul as a function of GS density
	}
	\label{fig:Rate}
	\vspace{-3mm}
\end{figure}

In \Fig{Rate} we plot the expected rate of the UAV backhaul as a function of the density of GSs, given an LTE backhaul. We can see that the expected rate initially increases as we increase the density of the GSs. However, due to the effect of interference the expected rate appears to behave differently depending on the UAV height. For the lowest UAV height the LOS probability on the interfering GSs is low due to building blockage, and therefore the rate monotonically increases with increased GS density. At the mid-range heights (\unit[60 and 80]{m}) the UAVs have a higher LOS probability on the interfering GSs; as a result increasing the density of the GSs improves the signal from the serving GS, but at the same time increases the aggregate interference. For the large heights (\unit[100 and 120]{m}) the UAV has a steep vertical angle to its serving GS, which results in a smaller area $\mathcal{W}$ illuminated by the UAV antenna and which limits interference. The height the UAVs will operate at will be largely determined by the end user link \cite{Galkin_20172}; however, an operator may wish to avoid operating the UAVs within the range of heights which cause deteriorated backhaul performance, if such an option exists. 

\begin{figure}[t!]
\centering
	\includegraphics[width=.45\textwidth]{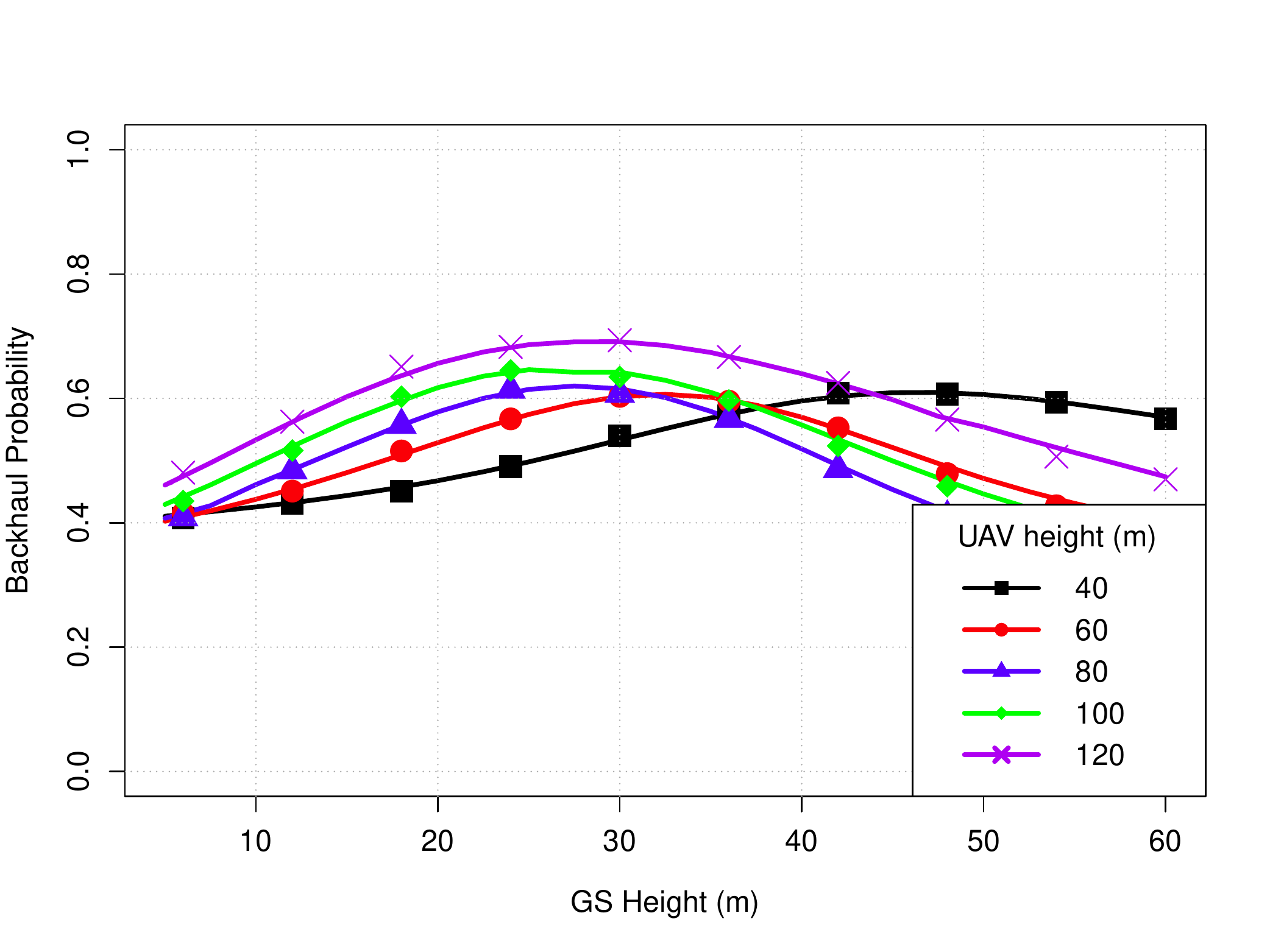}
    \vspace{-5mm}
	\caption{
	Backhaul probability for an LTE backhaul, given a GS density of \unit[1.25]{$/\text{km}^2$}
	}
	\label{fig:BackhaulBSHeight}
	\vspace{-3mm}
\end{figure}

In \Fig{BackhaulBSHeight} we consider the effect of the GS height on the backhaul probability of an LTE backhaul, for different UAV heights. We immediately observe that for larger GS heights the backhaul probability appears to deteriorate for all but the lowest UAV height. This effect is due to the increasing interference that is experienced by a typical UAV as the GS heights increase; the improved wireless channel to the serving GS does not compensate for the improved wireless channels to the interfering GSs. The GS height cutoff point above which the interference deteriorates appears to be around \unit[30]{m}, which corresponds to the height of LTE BSs in urban environments as proposed by the 3GPP model \cite{3GPP_2010}. These results suggest that when deploying dedicated GSs for backhauling the UAVs using the LTE band the operators should avoid placing the GSs any higher than the standard height used for the current terrestrial BS network, which also suggests that existing BS sites are suitable for hosting the backhaul GSs. It is also worth noting that the backhaul probability only marginally decreases for GS heights lower than \unit[30]{m}; this suggests that it is possible to provide UAV backhaul using GSs that are positioned at heights very close to ground level.

\begin{figure}[t!]
%\vspace{-8mm}
\centering
	\includegraphics[width=.45\textwidth]{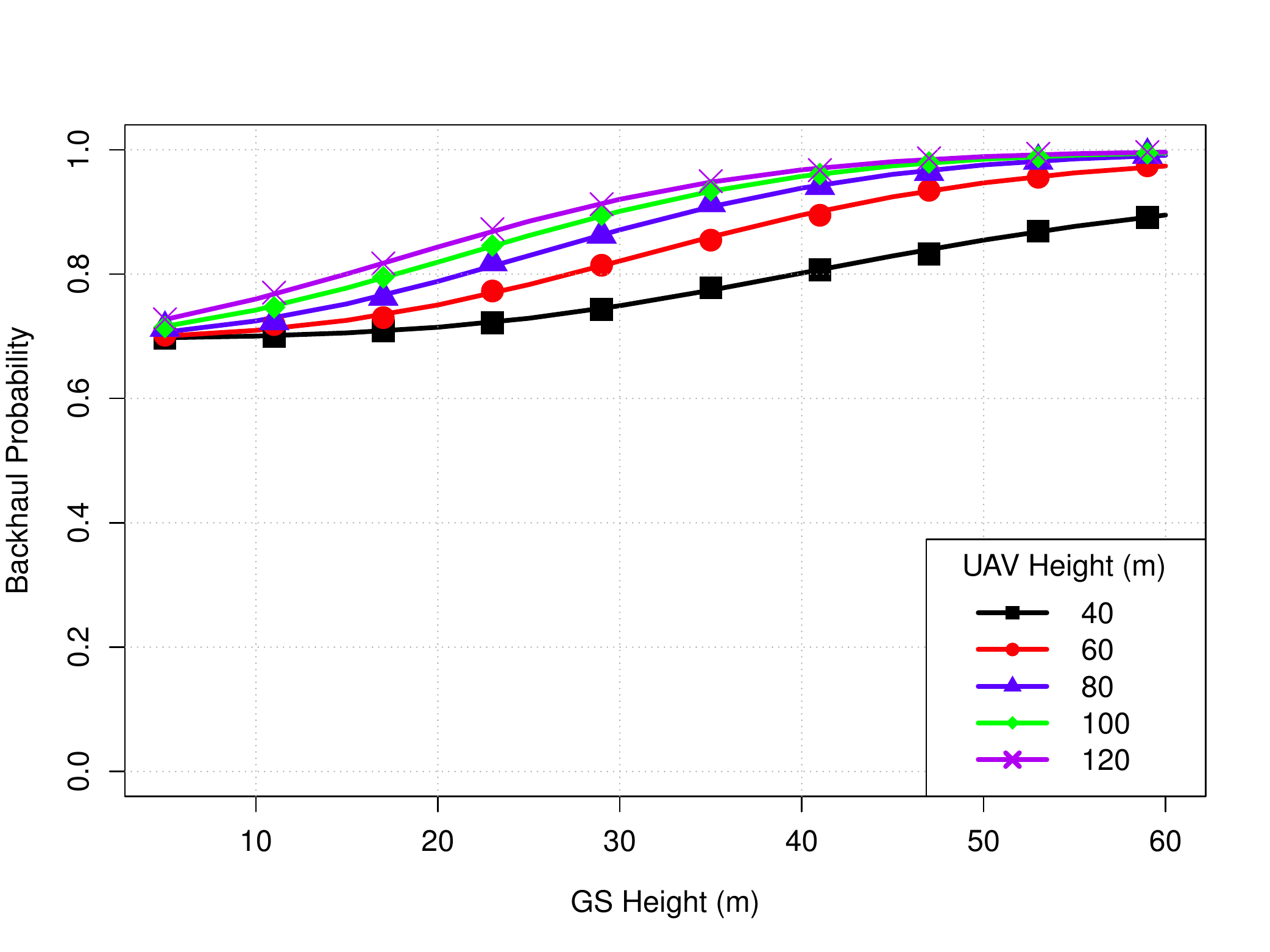}
	\vspace{-5mm}
	\caption{
	Coverage probability for a millimeter-wave backhaul, given a GS density of \unit[1.25]{$/\text{km}^2$}
%	\vspace{-30mm}
	}
	\label{fig:mmwaveGSHeight}
	\vspace{-5mm}
\end{figure}

In \Fig{mmwaveGSHeight} we show the effect of the GS height on the backhaul probability when the backhaul uses a millimeter-wave signal. We see that the backhaul probability monotonically increases with increased GS height. This is due to the increase in the LOS probability between the UAV and the serving GS. Recall that for a millimeter-wave signal high-directionality antennas are assumed on the part of both the UAV as well as the backhaul, as a result of this the UAV is assumed to receive no interference from other GSs, even when those GSs have LOS on the UAV. It follows then that the network operator should consider deploying the GSs as high as possible above the ground to maximise backhaul performance, which makes the existing BS sites sub-optimal for hosting the GS equipment, in contrast to the LTE backhaul case. As in the previous plots, we observe that greater UAV heights correspond to larger backhaul probability, showing the importance of operating UAVs at heights which can strike a balance between ensuring a good signal for the end user while simultaneously allowing the UAVs to meet their backhaul requirements.

\section{Conclusion}
In this paper we have used stochastic geometry to model a network of GSs that provide wireless backhaul to UAVs in an urban environment. Our model takes into account network parameters such as GS density and antenna characteristics, along with environmental parameters such as building density. We demonstrated that a good backhaul probability for the UAVs can be achieved with a GS density that is lower than the typical BS network density in an urban environment, and that LTE and millimeter-wave backhauls require different GS heights to maximise performance. 

In subsequent works we will consider more detailed models of the millimeter-wave channel which take into account factors such as shadowing, atmospheric and building signal absorption, as well as the impact of antenna misalignment.

\section*{Acknowledgements}
This material is based upon works supported by the Science Foundation
Ireland under Grants No. 10/IN.1/I3007 and 14/US/I3110. %B. Galkin, J. Kibi\l{}da, and L. DaSilva are with CONNECT, Trinity College Dublin, Ireland, email:  \{galkinb,kibildj,dasilval\}@tcd.ie.
\ifCLASSOPTIONcaptionsoff
  \newpage
\fi



\bibliographystyle{./bib/IEEEtran}
\bibliography{./bib/IEEEabrv,./bib/IEEEfull}
\end{document}